\documentclass[a4paper,12pt]{article}

\usepackage{bm}
\usepackage{graphics}
\usepackage[dvips]{graphicx}
\usepackage{latexsym,amsmath,amsfonts,amssymb}
\usepackage[latin1]{inputenc}
\usepackage[american]{babel}
\usepackage[dvips]{graphicx}
\usepackage{bbm}
\usepackage{subfigure}

\newcommand{\bbC}{\mathbb{C}}
\newcommand{\bbT}{\mathbb{T}}
\newcommand{\bbN}{\mathbb{N}}
\newcommand{\bbR}{\mathbb{R}}
\newcommand{\bbZ}{\mathbb{Z}}
\newcommand{\calR}{\mathcal{R}}
\newcommand{\calW}{\mathcal{W}}
\newcommand{\calJ}{\mathcal{J}}
\newcommand{\calT}{\mathcal{T}}
\newcommand{\calA}{\mathcal{A}}
\newcommand{\calC}{\mathcal{C}}
\newcommand{\calO}{\mathcal{O}}
\newcommand{\CP}{\mathbb{C}\mathrm{P}}
\newcommand{\dd}{\mathrm{d}}

\newcommand{\be}{\begin{equation}}
\newcommand{\ee}{\end{equation}}
\newcommand{\bea}{\begin{eqnarray}}
\newcommand{\eea}{\end{eqnarray}}

\begin{document}
\begin{titlepage}
\vspace{25pt}

\begin{center}
{\LARGE \bf Toric geometry and local Calabi-Yau varieties}
\end{center}
\vspace{10pt}

\begin{center}
{\Large \bf An introduction to toric geometry (for physicists)}

\end{center}

\vspace{15pt}

\begin{center}
{\large  Cyril Closset}\\

\vspace{25pt}
{\sl Physique Th\'eorique et Math\'ematique \\
and International Solvay Institutes \\
Universit\'e Libre de Bruxelles \\
CP 231, 1050 Bruxelles, Belgium}

\vspace{25pt}
{\tt  cyril.closset @ ulb.ac.be}

\end{center}

\vspace{20pt}

\begin{center}
\textbf{Abstract}
\end{center}
These lecture notes are an introduction to toric geometry. Particular focus is put on the description of toric local Calabi-Yau varieties,
such as needed in applications to the AdS/CFT correspondence in string theory. 

The point of view taken in these lectures is mostly algebro-geometric but no prior knowledge of algebraic geometry is assumed.
After introducing the necessary mathematical definitions, we discuss the construction of toric varieties as holomorphic quotients. We discuss
the resolution and deformation of toric Calabi-Yau singularities. We also explain the gauged linear sigma-model (GLSM) K\"ahler quotient construction.

These notes are based on lectures given by the author at the Modave Summer School in Mathematical Physics 2008.
\end{titlepage}

\tableofcontents

\section{Introduction}
The aim of these lectures is to explain toric geometry to young researchers in theoretical physics who might
have had no prior exposure to the basic concepts of algebraic geometry. Since the subject of algebraic geometry
is often seen as very abstract, on the one hand, while on the other hand string theorists routinely use it in a quite 
heuristic manner that might unsettle the young practitioner, I attempted here to find 
some kind of equilibrium between rigor and readability. We are lucky that toric geometry is precisely a very ``concrete'' area of algebraic geometry, so we can work on many examples.

An example that we will work with  extensively is a space called the \textit{conifold}, which has been studied from many perspectives by string theorists in the last two decades. The conifold began its physics career in the late '80s, in the context of the study of the large space of possible string compactifications \cite{Candelas:1988di,Candelas:1989js,Candelas:1989ug}.

The physicist approach taken here cannot replace the benefits of a formal algebraic geometry course, or of some good old-fashioned self-tuition in the mathematical literature, but my hope in preparing these lectures for the 2008 Modave Summer School was to at least make the subject look less scary to beginning graduate students. 

\subsection{A digression: supersymmetric gauge theories and D-branes at toric singularities}\label{recall SUSY gT}
A supersymmetric gauge theory in four dimensions is defined at the classical level by its Lagrangian density. For an introduction, see for instance \cite{Argurio:2003ym,Bilal:2001nv}. In superspace, consider the theory of  $k$ chiral superfields $\Phi_i$ ($i=1,\cdots, k$) charged under the gauge group in some representations. $\calW_{\alpha}$ is the gaugino chiral superfield, containing the field strength. We have
\be
\mathcal{L} = \int\mathrm{d}^2\theta\mathrm{d}^2\overline{\theta}\ \Phi^{\dagger}e^{2V}\Phi + \int\mathrm{d}^2\theta \{ \frac{\tau}{16\pi i} \mathrm{Tr} \mathcal{W}^{\alpha} \mathcal{W}_{\alpha} + W(\Phi)\}+ c.c.
\ee
where all flavor and gauge indices have been omitted. The superpotential $W$ is an holomorphic polynomial in the $\Phi^i$. The space of vacua is given by the vanishing of the $D$- and $F$-terms,
\bea
D^a &\equiv& \sum_i \Phi_i^{\dagger} T^a_{r_i} \Phi^i =0\, ,\label{D term constraint}\\ 
F_i^{\dagger} &\equiv& \frac{\partial W}{\partial \Phi^i}=0\, .
\eea
Here, the $\Phi^i$ are in arbitrary representations $r_i$ of the gauge group, so that the above conditions are matrix relations which transform non-trivially under the gauge group. It is an important result \cite{Luty:1995sd} that the full space of vacua can be described as an \textit{algebraic variety}, which roughly means that it is a complex hypersurface (or intersection thereof) in $\bbC^m\cong \{x_1,\cdots,x_m\}$. Here the $x_j$ are gauge invariant polynomials, for instance corresponding to operators of the form
\be
x \sim  \mathrm{Tr}\, P(\Phi),
\ee
with $P$ a polynomial function of the $\Phi^i$, and the trace is over all gauge indices. Restricting to gauge invariant polynomials is equivalent to fixing the gauge freedom through the imposition of the D-terms constraints (\ref{D term constraint}), as was rigorously shown in \cite{Luty:1995sd}. The advantage is that we are now dealing with holomorphic quantities (only $\Phi$ appears, not $\Phi^{\dagger}$). The relations $F_i=0$ then imply relations between the variables $x_j$, which can be written as polynomial relations
\be
p_1(x_1,\cdots,x_m)=0,\quad p_2(x_1,\cdots,x_m)=0, \cdots
\ee
This is precisely the way to define an algebraic variety. We will make this more precise in section \ref{section Affine varieties}.

From this discussion, we could guess that an algebro-geometric language can be very useful in order to deal with supersymmetric theories. See \cite{Gray:2008yu} for a recent work emphasizing this basic point: the space of  vacua of any supersymmetric gauge theory is
\textit{algebraic} in nature.

Where does toric geometry fit in this context? There exist a very interesting class of supersymmetric gauge theories whose space of vacua is \textit{toric}. These theories are the so-called \textit{toric quiver gauge theories}. They appear naturally in string theory as the low energy effective theory of D3-branes probing  toric Calabi-Yau singularities.

If you have no idea what the previous paragraph refers to, do not panic. The purpose of these lectures is precisely to explain what a ``toric Calabi-Yau singularity'' is, and to offer some basic mathematical tools necessary to deal with models involving them in string theory.

These lectures can also serve as a starting point to learn more about the use of toric geometry in many other areas of string theory.
For instance, one can describe many Calabi-Yau compactification manifolds as hypersurfaces in compact toric varieties, as reviewed in \cite{Greene:1996cy, Kreuzer:2006ax, Bouchard:2007ik, Denef:2008wq}. We will not review this construction here, focusing instead on local properties. We nevertheless explain the general case of compact toric varieties, emphasizing the importance of the ``local'' toric affine varieties as building blocks.

\subsection{Outline of the lectures}
Since toric geometry is a part of algebraic geometry, we will start in the next section with an introduction to the basic concepts of algebraic geometry. We will first define affine varieties, explaining how there is an equivalence between geometric objects (the varieties) and algebraic objects (some particular sets of polynomials called prime ideals). Some important definitions are relegated to the Appendix.
Next we will briefly talk about projective space, as a warm up, since it is a simple example of a toric variety.

In section \ref{section CY condutin} we will discuss the Calabi-Yau condition. To do so we will need to introduce the notion of a line bundle. That part of the lectures is not self-contained. It uses differential geometry concepts that are hopefully familiar to the general reader, mostly the basics of the theory of fiber bundles. 

In section \ref{Section TG1} we will delve into the core of the subject, defining toric varieties as  particular holomorphic quotients, and showing how to introduce local coordinates in term of affine varieties (affine patches). Remark that we will mainly be interested in \textit{local} properties, and so we will mostly concentrate on non-compact toric varieties. In particular we will consider Calabi-Yau toric varieties, which are always non-compact.

In section \ref{Section: singularities} we will introduce the notion of singularity in algebraic geometry, and we will show how we can deal with singular points in the toric case.

In section \ref{section GLSM} we introduce a second way to define toric varieties, the K\"ahler quotient, also known as gauged linear sigma-model.\newline

I tried to make the following as self-contained as I could, but general knowledge of complex geometry might help at times, especially in section \ref{section CY condutin}, as already mentioned.
Good introductions to complex geometry and Calabi-Yau manifolds can be found for instance in \cite{Greene:1996cy, Bouchard:2007ik, Hubsch:1992nu}. The more thirsty student might plunge into \cite{joyce}, which I found a very good and surprisingly physicist-friendly mathematical reference.

I also hope that these lectures will serve as entry point into the mathematical literature on toric geometry, such as \cite{Fulton}. Posterior developments as explained in \cite{Cox} are also important, as they actually simplify matters. See \cite{Skarke:1998yk} for a nice account from a physicist perspective.

\section{Algebraic geometry: the gist of it}

We know that in geometry we always deal with some bunch of ``points'' that has more or less structure to it.
A set of points together with a topology is called a topological space. Recall that a topology is what you define to be the open sets in your space, 
hence it provides a notion of locality.
A topological space that locally looks like the euclidian space $\bbR^n$ is called a manifold. If moreover the transition functions are differentiable ($C^{\infty}$ for instance), it is called a differentiable manifold.

Smooth algebraic varieties can be seen as particular kind of manifolds which are simpler in some sense. Roughly  speaking, they can be thought of as manifolds with rational transition functions\footnote{For toric varieties we will see that it is precisely that.}. On the other hand, generic algebraic varieties are \textit{not} manifolds, since they allow for various singularities; in that sense they are more general.

Remark that it is possible to define algebraic varieties \textit{intrinsically}, in a way similar to what one does in differential geometry, but for doing so we would need to introduce the language of sheaves, and that would carry us too far afield.
We will follow the more down to earth route, which defines algebraic varieties \textit{extrinsically} as the algebraic set of zeros of some polynomials.
Given a function $f: \bbR^n\rightarrow \bbR$, we can define a subset of $\bbR^n$,
\begin{displaymath}
\bbR^n \supset \Sigma \quad =\quad \{f^{-1}(0)\} \quad=\quad \{x\in \bbR^n\, |\,  f(x)=0\},
\end{displaymath}
which locally inherits its manifold structure from $\bbR^n$. However, this $\Sigma$ is badly singular in general. If we restrict $f$ to be a polynomial,  things become much more tractable. It is one of the great advantages of the algebraic side of algebraic-geometry that singularities become easier to deal with.

Therefore we are now considering algebraic equations only. Hence it is very convenient to work with polynomials valued in $\bbC$, because $\bbC$ is algebraically complete. From now on, unless otherwise stated, all variables are $\bbC$-valued, and by dimension we always mean complex dimension (half the real dimension).

In this section we will first define affine varieties, which are the basic objects of algebraic geometry.
Some algebraic definitions are reviewed in Appendix \ref{appendixAlgebra}. Next we define the projective space $\CP^n$, which provides us with a particular example of the holomorphic quotient construction that we will encounter 
in detail when we define toric varieties in section \ref{Section TG1}. For completeness we also define projective varieties, which are subvarieties of $\CP^n$.

\subsection{Affine varieties}\label{section Affine varieties}
Varieties defined as algebraic subset of  $\bbC^n$ lead to the concept of \textit{affine varieties}.
Consider $\bbC^n= \{(x_1,\cdots,x_n) \}$. Associated to it, we have the ring of polynomials in $n$ variables, which is denoted by
\be
\calR_n \equiv \bbC[x_1,\cdots,x_n]\, .
\ee
It is obviously a ring (it is an additive group together with an associative product, distributive with respect to the addition); moreover it is a commutative ring. An \textit{algebraic subset} $Z(\calT)$ of $\bbC^n$ is defined as the zero locus of a set of polynomials $\calT \subset \calR_n$:
\be\label{algSubsetZ}
Z(\calT) = \{(x_1,\cdots, x_n)\in \bbC^n \,|\,  p_i(x_1,\cdots,x_n)=0, \forall p_i \in \calT  \}.
\ee
On the other hand, for any subset $Y\subset \bbC^n$, we denote the set of all polynomials that vanish on $Y$ by $\calJ(Y)$. A natural question to ask 
is what is the relation between $\calJ(Z(\calT))$ and $\calT$. This is the content of the famous Hilbert's Nullstellensatz. See the Appendix \ref{appendixAlgebra}.

The whole idea of algebraic geometry is that you can define a space by the algebra of functions defined on it%
\footnote{Note that the ring of polynomials is naturally an \textit{algebra} too.}.
Let us look at the polynomials which give rise to well defined functions on the algebraic set (\ref{algSubsetZ}). Two polynomials $p_1$ and $p_2$ will take the same value on $Z(\calT)$ if $p_1-p_2=t$, with some $t\in \calT$, since $t$ vanishes on $Z(\calT)$ by definition. We then only need to consider
the equivalence classes of polynomials in $\calR_n$ that are linearly equivalent up to elements of $\calT$. This is denoted by
\be\label{Asigma}
\calA(Z(\calT)) = \bbC[x_1,\cdots,x_n]/\calT.
\ee
We want this quotient to define a proper ring of functions on $Z(\calT)$. This happens if $\calT$ is an \textit{ideal}
of the ring $\bbC[x_1,\cdots,x_n]$. An ideal of a ring $R$ is a subset $I\subset R$ such that $I$ is a subgroup for the addition
and  is invariant under multiplication by any element in $R$. Given any set of polynomials, it is not difficult to extend it into
a full-fledged ideal, as one can see in the examples below. One usually denote the ideal generated this way by $(p_1,\cdots,p_k)$.

\paragraph{Examples:} 
\begin{itemize}
\item Take the ring $\bbC[x]$ of polynomials in $x$. The set $\{x\}$ is not an ideal (for instance it is not even a subgroup), but we can generate one simply by multiplying with every element of $\bbC[x]$. The ideal, denoted $(x)$, is simply the set of all polynomials without constant term. The quotient by the ideal
simply gives the constants:
\be
\bbC[x]/(x) = \bbC.
\ee
\item Consider  the ideal $(x^2)$ instead. The quotient $\bbC[x]/(x^2)$ is a ring generated by the two elements $\{1,x\}$ such that $x.x=0$. 
Such a $x$ is called a zero divisor. 
\item On the ring $\bbC[x,y]$, consider the ideal $(xy)$. The quotient ring $\bbC[x,y]/(xy)$ has two zero divisors ($x$ and $y$). 
\end{itemize}

This last example corresponds to the surface $xy=0$ in $\bbC^2$. It consists of two branches which meet at the origin.
In general, any algebraic set will consist of several ``branches'',
\be\label{decomIrreCOm}
Z(\calT)= \Sigma_1 \cup \cdots \cup \Sigma_m,
\ee
and correspondingly the quotient ring  (\ref{Asigma}) will have zero divisors. 
To avoid zero divisors, one must ask that the ideal be \textit{prime} (see the Appendix \ref{appendixAlgebra} for the definition).
In our example, $(xy)$ is not prime, but it has a decomposition in two prime factors $(x)$ and $(y)$. These two
ideals correspond to the two ``branches'' $x=0$ and $y=0$.

Each component in the decomposition (\ref{decomIrreCOm}) is called \textit{irreducible} if it cannot be decomposed further.
\newline

\textbf{Definition:} An  \textit{affine variety} is an irreducible algebraic subset of $\bbC^n$.
\newline

It is called ``affine'' simply because it is defined in $\bbC^n$, which is an affine space (i.e. a vector space where you can shift the origin anywhere).

The very important thing to remember is that there is a one-to-one correspondence between affine varieties and prime ideals:
\be
\Sigma= Z(P)\quad  \stackrel{1-1}{\longleftrightarrow} \quad \calA(\Sigma)= \bbC[x_1,\cdots,x_n]/P.
\ee
 This is a consequence of the Hilbert's Nullstellensatz, which implies that if $\calT=P$ is a prime ideal%
\footnote{Actually this holds for $P$ \textit{radical}, which is a weaker condition. The one-to-one correspondence is between algebraic sets and radical ideals. Remark that in dimension one, it implies that a polynomial with isolated zeros is fully determined
by its roots; the Nullstellensatz is a generalisation of the fundamental theorem of algebra to higher dimensions.}
, then the set of polynomials vanishing on $Z(P)$ is $P$ itself:
\be 
\calJ(Z(P))=P.
\ee

\textbf{Definition:} The ring $\calA(\Sigma)$ defined as in (\ref{Asigma}),
\be
\calA(\Sigma) = \bbC[x_1,\cdots,x_n]/P,
\ee
is called the \textit{coordinate ring}, or \textit{structure ring}, 
of the affine variety $\Sigma$. This construction is  familiar from supersymmetric theories, as recalled in section \ref{recall SUSY gT}: there the $x_i$ are the gauge invariants operators,
and $P$ is generated by the F-terms. The structure ring in that case is called the \textit{chiral ring}.

\paragraph{Example: the conifold.} The ubiquitous conifold, $\calC_0$, which has been such a central tool in recent developments in string theory, is 
an affine variety defined by a single equation in $\bbC^4$,
\be\label{focofCOniiii}
x_1x_2-x_3x_4=0.
\ee
Mathematicians call it a ``threefold ordinary double point'', or \textit{node}.
Its coordinate ring is
\be
\calA(\calC_0)=  \bbC[x_1,x_2,x_3,x_4]/ (x_1x_2-x_3x_4).
\ee

\subsection{Projective varieties}\label{subsect: Proj varieties}
Affine varieties, being defined by polynomial equations in $\bbC^n$, are not compact. The projective space $\CP^n$ is the simplest example 
of a compact algebraic variety (actually it is toric too).  The standard way to define it is as the set of complex lines in $\bbC^{n+1}$,
\be
\CP^n = \frac{(\bbC^{n+1}\backslash \{0\})}{\bbC^*}.
\ee
The action of $\bbC^*=\bbC\backslash \{0\}$ is to multiply all coordinates in $\bbC^{n+1}$ by $\lambda\in \bbC^*$, which defines the equivalence relation
\be\label{Proj identification}
[x_0,\cdots,x_n] \sim [\lambda x_0,\cdots,\lambda x_n].
\ee
The origin $\{0\}$ was removed before taking the quotient so that $\bbC^*$ may act freely. The resulting space is fully regular.
The $x_i$ are called homogeneous coordinates, and a point in $\CP^n$ is represented by the equivalence class $[x_0,\cdots,x_n]$.
We can cover $\CP^n$ with $n+1$ affine patches, one for each $x_i\neq 0$. 
The local coordinates on the $i$-patch are $z^{(i)}_{k}=x_k/x_i$, and
the transition functions are  the rational functions
\be
z_k^{(i)}(z^{(j)})= \frac{z_k^{(j)}}{z_i^{(j)}}\, .
\ee
The Riemann sphere $\CP^1$ is the best known example. It has two patches, and the transition function on the equator is $z_N=1/z_S$.
\newline
We can define subvarieties of $\CP^n$ by taking the vanishing locus of a set of polynomials $p_i \in \calR_{n+1}$.
For the equations $p_i=0$ to make sense, they should be constant on any equivalence class $[x_0,\cdots,x_i]$, which
means the $p$'s are \textit{homogeneous} (i.e. they are sums of monomials of fixed degree):
\be
p_i(x_0,\cdots,x_n) \sim \lambda^d p_i(x_0,\cdots,x_n).
\ee

\textbf{Definition: } Given a homogeneous prime ideal $P_h$ in $\calR_{n+1}$,  the associated \textit{projective variety} is defined as
\be
\Sigma(P_h) = \{\, [x_0,\cdots,x_n] \quad | \quad p_i=0 \quad \forall p_i\in P_h \subset \calR_{n+1} \}. 
\ee
It is easy to check that if the $p_i$'s are homogenous of degree $d$, so is the ideal $(p_i)$.

The \textit{homogeneous coordinate ring} is denoted by
\be
\mathcal{S}(\Sigma) = \calR_{n+1}/P_h \, .
\ee

\textbf{Projective plane curves.}
In $\CP^2$, consider a hypersurface defined by a single polynomial $p$  of degree $d$. If moreover
\be
\frac{\partial p(x)}{\partial x_i} = 0 \, \forall i, \quad \forall x \quad \mathrm{s.th.} \quad p(x)=0\, ,
\ee
the curve is regular; it is a Riemann surface. Such Riemann surfaces are classified by their genus. There exists a theorem stating that
\be
g= \frac{(d-1)(d-2)}{2}\, .
\ee
In particular, for $d=3$, we have a \textit{torus}, or elliptic curve ($g=1$). The general equation reads
\be
\sum_{i+j+k=3} c_{ijk}\, x_0^ix_1^jx_2^k =0.
\ee
We have 10 parameters here. However 9 of them can be removed by a $Gl(3,\bbC)$ transformation on the homogeneous coordinates.
This leaves us with one parameter, which is  basically the complex structure modulus of the torus. We will come back
to the important issue of complex structure moduli later on in these lectures.\newline

Remark that there are many more algebraic varieties than just affine and projective ones. 
In general, one can patch together affine varieties to obtain any algebraic variety, similarly to the idea of patching together open
sets to form manifolds in differential geometry.
We will see this explicitly in the simpler context of toric varieties.

\subsection{Spectrum and scheme, in two words}
Let us introduce the notion of spectrum of a ring. This is done only to set a useful notation that you might 
often encounter in the literature.
The concepts of spectrum and scheme stem from taking seriously
the idea that it is really the algebra of functions on it which defines a space. One starts with a purely algebraic object : given
\textit{any} ring $A$, one defines its \textit{spectrum}
\be
\mathrm{Spec} (A) \equiv \{P\subset A\},
\ee
to be the set of all prime ideals of $A$ (except $A$ itself). This set can be given a natural \textit{topology},
and it is then shown that, in the particular case of the coordinate ring of an affine 
variety,
\be\label{def of Spectrum}
\mathrm{Spec} (\calA(\Sigma)) \cong \Sigma,
\ee
up to important subtleties that we shall willfully skip (in particular we are really talking about the maximal ideals here).
The scheme structure is then obtained by introducing local coordinates by means of a so-called \textit{structure sheaf} (for interesting introductions to sheaf concepts in physics, see for instance \cite{Sharpe:2003dr,Aspinwall:2004jr}).

\section{The Calabi-Yau condition}\label{section CY condutin}
For applications to ``physics'' (string theory in fact), we are mostly dealing with so-called Calabi-Yau (CY) varieties.
The Calabi condition is a topological condition that implies (by Yau's theorem) that there exist a Ricci-flat metric on
the variety which satisfies that condition. Hence a Calabi-Yau manifold is a vacuum solution to the Einstein field equations,
which is a necessary condition for being a semi-classical background of string theory (in the absence of flux). 

The extension to Calabi-Yau varieties with singularities is interesting too, 
because many new stringy phenomena like topology changing processes
occur in the presence of singularities (since in the -inadequate- language of Riemannian geometry, one could say that a singular point has infinite curvature, hence Planck-scale effects must dominate there). Algebraic geometry offers some tools to tackle these important string theory questions.

Moreover, in the context of $AdS$/CFT, one considers objects called D-branes located at singular points in local Calabi-Yau varieties.
There is an interesting correspondence between the algebraic structure of the singularity and the details of the conformal field theory:  the number of gauge groups, the matter content and the classical interactions in the CFT can in principle be deduced from the geometry alone.

In this section we consider algebraic \textit{manifolds}, i.e. \textit{non-singular} algebraic varieties.
It is fair to warn the reader that we will be applying results of this section in singular cases in the next section,
so keep your eyes peeled.

An algebraic manifold is obviously a complex manifold: all the quantities we are dealing with are holomorphic by construction,
and the variety inherits its complex structure from the embedding space $\bbC^n$ or $\CP^n$.

In this section, since we deal with manifolds, we can take a more direct, ``intrinsic'', differential-geometric standpoint.
This will simplify matter, since differential geometry is bound to be more familiar to the reader. 

\subsection{Holomorphic vector bundles and line bundles}
Consider a complex manifold $X$ of dimension $m$. On every open set we have local coordinate functions $z_1,\cdots, z_m$,
and we can define the exterior algebra of these coordinate functions, generated by \textit{one-forms}
\be
\dd z_1, \cdots, \dd z_m .
\ee
At any point $p$ in the open set, $\{\dd z_i(p)\}$ form a basis for the \textit{holomorphic cotangent space} $T^*_p X$ at $p$.

The multiplication operation on forms is the exterior product. All in all we have $2^m$ linearly independent elements
\be
1, \quad \dd z_i, \quad \dd z_{i_1}\wedge \dd z_{i_2},\quad  \cdots,\quad \dd z_{i_1}\wedge\cdots\wedge \dd z_{i_m} \, ,
\ee
which form a graded algebra. At each degree, $p$-forms at any particular point span a vector space of dimension $\frac{n!}{p!(n-p!)}$.

Using holomorphic $Gl(m,\bbC)$-valued transition functions, we can patch all cotangent spaces together
into the \textit{holomorphic cotangent bundle} $T^*X$:
\be
\bbC^m \quad \longrightarrow\quad  T^*X \quad  \stackrel{\pi}{\longrightarrow}\quad  X \, ,
\ee
which is itself a manifold of dimension $2m$. This is a particular case of an holomorphic vector bundle $E$, 
\be
\bbC^k \quad \longrightarrow\quad  E \quad  \stackrel{\pi}{\longrightarrow}\quad  X \, ,
\ee
with $\bbC^k$ the fiber, and $\pi$ the natural projection, which is an holomorphic map. $k$ is called the rank of the bundle.

\textbf{Definition:} An \textit{holomorphic line bundle} (or \textit{line bundle} for short) is an holomorphic vector bundle of rank one.

A very important line bundle is the \textit{canonical bundle} $K_X$. It is defined as the $m^{\mathrm{th}}$ exterior product of $T^*X$,
\be
\bbC \quad \longrightarrow\quad  K_X \equiv \Lambda^{(m,0)}T^*X \quad  \stackrel{\pi}{\longrightarrow}\quad  X \, .
\ee
Sections of the canonical bundle are holomorphic m-forms, that we can write (on each coordinate patch)
\be\label{sectionsCanLB}
\Omega = f(z) \dd z_1\wedge\cdots\wedge\dd z_m,
\ee
for $f(z)$ some holomorphic function.

\subsection{Calabi-Yau manifolds. K\"ahler and complex moduli}
The Calabi-Yau condition is that the canonical bundle be \textit{trivial}, i.e.
\be
\Lambda^{(m,0)}T^*X \quad \cong \quad \bbC\, \times \, X\, .
\ee
This implies the existence of a never vanishing  \textit{global section}. Standard arguments then imply that
the function $f(z)$ in (\ref{sectionsCanLB}) must be a constant. This unique (up to rescaling by a constant) $\Omega$ is usually called the 
holomorphic $m$-form of the Calabi-Yau manifold $X$.

\paragraph{K\"ahler structure.} A complex manifold can be endowed with a \textit{K\"ahler structure}. There is no room
here to explain in detail what this is, see \cite{Greene:1996cy,Bouchard:2007ik}.
In two words though, a K\"ahler structure is a symplectic structure compatible with the complex structure:
you need a closed and non-degenerate $(1,1)$-form $\omega$. The nice thing is that complex structure plus K\"ahler structure implies 
there is a compatible Riemannian structure, i.e. a \textit{hermitian metric}. This metric is defined by
\be\label{kahlermetric} 
g( \partial_z,\bar{\partial}_{\bar{z}}) = \omega(\partial_z, i\, \bar{\partial}_{\bar{z}})
\ee
for any two vectors $\partial_z$, $\bar{\partial}_{\bar{z}}$ in the tangent space (holomorphic and anti-holomorphic).

The Kahler form $\omega$ is a representative of a Dolbeault cohomology class%
\footnote{Again a word I will not define. See for instance \cite{Greene:1996cy}.}
\be
[\omega] \, \in \, H^{1,1}(X).
\ee
$[\omega]$ is called the K\"ahler class of $\omega$. 

Now, we can state Yau's theorem (Yau proved a conjecture made earlier by Calabi): 
\paragraph{CY Theorem :} Given $X$ a compact complex manifold \textit{with trivial canonical bundle}, and given
a K\"ahler form $\tilde{\omega}$ on $X$, there exist a unique Ricci flat metric in the K\"ahler class of $\tilde{\omega}$.
That is, a unique Ricci-flat metric given by (\ref{kahlermetric}) for some $\omega\in [\tilde{\omega}]$.

On the other hand, it is ``easy'' to show that Ricci-flatness implies the triviality of the line bundle. 
For a non-compact manifold, the theorem does not hold (strictly speaking). One can still find a Ricci-flat metric in general, but one must specify some boundary conditions at infinity.

\paragraph{K\"ahler moduli space.} Given a Calabi-Yau manifold $X$, we see there are continuous families of Ricci-flat metrics, one for each cohomology class
\be
[\omega]\quad =\quad \sum_{i=1}^{h^{1,1}} \lambda_i \, [\omega]^i \, .
\ee
These parameters $\lambda$ are coordinates in a vector space $H^{1,1}(X)$ (here the $[\omega]^i$ are basis vectors). It is called the \textit{K\"ahler moduli space} of $X$. Its dimension is denoted by $h^{1,1}$.

\paragraph{Complex moduli space.} Given an algebraic variety, if one modifies the equation continuously, varying some parameters, the variety
will be ``deformed'' accordingly. This is called a \textit{variation of the complex structure}. 

Consider the example of the torus of section
\ref{subsect: Proj varieties}; we saw there are 10 parameters one can vary, but 9 of them do not change the complex structure, because they are 
just a linear reshuffling of the embedding space coordinates, so the complex moduli space of the torus is one dimensional.

Consider also the conifold, defined by $x_1x_2-x_3x_4$. If I write, for instance,
\be
x_1x_2-bx_3x_4+ c x_4=0,
\ee
the constants $b$ and $c$ can obviously be absorbed in a redefinition of $x_3$, with $x_3'=bx_3-c$. For an affine variety in $\bbC^n$,
we can transform the variables by 
\be
Gl(n,\bbC) \ltimes T_n,
\ee
with $T_n$ the group of translations. In the case of $\calC_0$ in $\bbC^4$, we have 15 possible parameters for a generic polynomial of degree 2.
However most of them can be removed by a $Gl(4,\bbC) \ltimes T_4$ transformation. One can check that the only parameter which cannot be removed by such a transformation is the constant term,
\be
x_1x_2-x_3x_4-a=0.
\ee
Such a space is called the \textit{deformed conifold}, and it is regular.

The space of all complex deformations of an algebraic variety $X$ is called the \textit{complex moduli space} of $X$.
It is a rather complicated space. Its \textit{linearisation} (the tangent space) is given
by the cohomology group $H^{m-1,1}(X)$ ($m$ the dimension of $X$) in the case of Calabi-Yau manifolds. 
In general, the question is much more complicated. In the particular case of the theory of complex deformations of toric
Calabi-Yau singularities, there is some important results to be learned, as we will see.

\subsection{Divisors and line bundles}
\textbf{Definition: } A (Weyl) \textit{divisor} $D$ of a complex variety $X$ is a linear combination 
(a formal sum with integer coefficients) of codimension one irreducible subvarieties,
\be\label{defDivisor}
D = \sum_i n_i V_i, \quad n_i\in \bbZ, \quad V_i \subset X.
\ee
If all $n_i \geq 0$, the divisor $D$ is said to be \textit{effective}.

To any line bundle $L$ with a \textit{regular section} $s$ 
(which means that on any open set $U_{\alpha}$, $s_{\alpha}$ is a polynomial in the local coordinates)
we have an associated hypersurface $Y$ in $X$ defined by
\be
Y = \{ s(p)=0, \quad p\in X \}.
\ee
We can decompose $Y$ into irreducible parts. On any affine patch, the polynomial $s_{\alpha}$ can be factorized in $\bbC[x_1,\cdots, x_n]$.
In fact, $(s_{\alpha})$ is decomposed into prime ideals, and one keeps track of the multiplicity%
\footnote{There is a multiplicity because the ideal $(s_{\alpha})$ is not radical in general.}
 $n_i$ of each distinct ideal $P_i$.
The prime ideal $P_i$ corresponds to the subvariety $V_i$ in (\ref{defDivisor}). More precisely, one should of course 
patch all the $V_{i}^{\alpha}$ together to construct $V_i\subset X$.

Going the other way around, an effective divisor $D= \sum_i n_iV_i$ defines a line bundle, denoted $\calO_X(D)$. 
By definition its sections will vanish on each $V_i$ with a zero of order $n_i$.

On can generalize this construction to any divisor, where now $n_i<0$ corresponds to a pole of order $n_i$ for the corresponding sections
of $\calO_X(D)$.
\newline

\textbf{Example.} On $X=\CP^n$, we can set $z_i=0$ ($z_i$ an homogeneous coordinate). It corresponds to the hyperplane $H$
(any $H_i= \{z_i=0\}$ is linearly equivalent to the others). A general divisor is then $D=nH$, $n\in \bbZ$. Its associated
line bundle is usually denoted $\calO(n)$.
Note that $\calO(-1)$, corresponding to $D=-H$, is really the dual of the hyperplane line bundle (i.e. its sections are in $\mathrm{Hom}(\calO(1), \bbC)$).
It is called the tautological line bundle of $\CP^n$.

\section{Toric geometry 1: The algebraic story}\label{Section TG1}
We are now ready to discuss toric geometry. In this section we define a toric variety as a particular \textit{holomorphic quotient} of $\bbC^n$.

\textbf{Definition: } A \textit{toric variety} $X$ (of dimension $m$) is an algebraic variety containing the algebraic torus $\bbT=(\bbC^*)^m$
as a dense open subset, together with a natural action $\bbT\times X \rightarrow X$.

We can write $X$ as
\be\label{ConstruTV1}
X_{\Delta} = \frac{\{ \bbC^{n}\backslash Z_{\Delta} \}}{G}.
\ee
Here, the group
\be
G \cong (\bbC^*)^{n-m} \times \Gamma,
\ee
is an algebraic torus times an abelian discrete group $\Gamma$. 
This construction generalizes the one for projective spaces. For it to make sense, we have to specify a set of points $Z_{\Delta}\subset \bbC^n$,
 and of course we must know how $G$ acts on $\bbC^n$.

\subsection{Cones and fan. Homogeneous coordinates}
All this data defining a toric variety can be encoded in a simple auxiliary object called a \textit{fan}. Hence the fan can be taken to define 
the toric variety. An equivalent definition will be in term of the gauged linear sigma-model of section \ref{section GLSM}: the
same data  is present in both definitions, in particular the charge matrix to be defined momentarily. Moreover, this data is \textit{combinatoric}, which means that is is given by discrete quantities. What makes toric geometry attractive is that complicated geometric problems can often be reduced to simpler combinatoric problems.

Let $N\cong \bbZ^m$ be a lattice, and $N_{\bbR}= N \otimes \bbR$ the vector space obtained by allowing real coefficients.
\newline

\textbf{Definition: } A strongly convex rational polyhedral cone $\sigma\subset N_{\bbR}$, or \textit{cone} for short, is a set
\be
\sigma = \{\sum_i a_i v_i \quad | \quad a_i \geq 0  \},
\ee
generated by a finite set of vectors $\{v_i\}_{i=1}^{n}$ in $N$, and such that $\sigma \cap (-\sigma)=\{0\}$ (``strong convexity'').
\newline

\textbf{Definition:} A \textit{fan} is a collection $\Delta$ of cones in $N_{\bbR}$ such that \newline
(i) each face of a cone is also a cone,\newline (ii) the intersection of two cones is a face of each.
\newline

Let us call $\Delta(1)$ the set of one-dimensional cones in $N_{\bbR}$.
The corresponding vectors in $N$ are denoted $(v_1, \cdots, v_n)$.
To each $v_i$, one associates a \textit{homogeneous coordinate} $z_i$.
These are the coordinates on $\bbC^n$ in the holomorphic quotient construction (\ref{ConstruTV1}).

Remark that we always have $n\geq m$. The $(m\times n)$ matrix
\be\label{LinearCnCm}
(v^k_i)=(v_1^k, \cdots, v_n^k)
\ee
(with $k=1,\cdots, m$) induces a map 
\be
\phi: \bbC^n\rightarrow \bbC^m : (z_1,\cdots, z_n) \mapsto (\prod{_{i=1}^n}z_i^{v_i^1}, \cdots, \prod{_{i=1}^n}z_i^{v_i^m}).
\ee
We define $\tilde{G}=(\bbC^*)^{n-m} \subset G$ to be the kernel of $\phi$:
\be
\tilde{G} = \mathrm{Ker(\phi)}. 
\ee
It is easily seen that $\tilde{G}$ acts on $\bbC^n$ as 
\be\label{action of Gtilde}
\tilde{G} \supset (\bbC^*)_a\, :\quad (z_1,\cdots, z_n) \mapsto (\lambda^{Q^a_1}z_1, \cdots ,\lambda^{Q^a_n}z_n)
\ee
for each $a$, where the  charge vectors $Q^{a}$ are in the kernel of the linear map (\ref{LinearCnCm}), that is:
\be\label{QinkerV}
\sum_i (v^k_i)Q^a_i=0.
\ee
Hence, practically speaking, given a fan with $n$ vectors in $N$ we must find the $n-m$ linear relations among them. The coefficients
are precisely the $Q^a_i$ above.

The discrete group $\Gamma\subset G$ is defined as
\be
\Gamma = N/N' \, ,
\ee
where $N'\subset N$ is the sublattice generated over $\bbZ$ by the vectors $v_i$. The quotient by this $\Gamma$ gives rise to so-called \textit{orbifold} singularities.

Last but not least piece of data in the construction, the zero set $Z_{\Delta}$ is found as follows: For any subset of $\Delta(1)$ 
(corresponding to vectors $v_{i_1},\cdots, v_{i_l}$) which do \textit{not}
generate a cone in $\Delta$,  associate an algebraic set $V_{i_1,\cdots,i_l}$ defined by $z_{i_1}=\cdots= z_{i_l}=0$. 
Then $Z_{\Delta}$ is the union of all these subsets of $\bbC^n$.
\newline

We'd better move on to examples.\newline
\begin{figure}[t]
\begin{center}
\includegraphics[height=5cm]{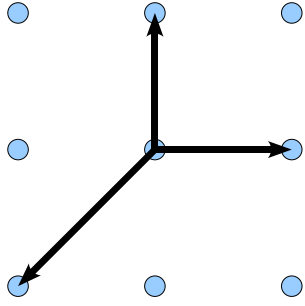}
\caption{\small The toric fan of $\CP^2$.  Notice that it contains 7 cones: three top-dimensional ones (of dimension 2), three 1-dimensional ones (generated by the vectors), and a single 0-dimensional one (the point in the center).
\label{fanCP2}}
\end{center}
\end{figure}
\begin{itemize}
\item  Consider the fan  in figure \ref{fanCP2}, generated by 3 vectors in $N\cong\bbZ^2$:
\be
v_1=(1,0), \quad v_2=(0,1), \quad v_3=(-1,-1).
\ee
The one relation $v_1+v_2+v_3=0$ gives a single charge vector (see (\ref{QinkerV}))
\be
Q= (1,1,1),
\ee
so we have the following group action of $G=\bbC^*$ on the homogeneous coordinates:
\be
G: (z_1,z_2,z_3) \mapsto (\lambda z_1,\lambda z_2,\lambda z_3).
\ee
Moreover, one sees that $Z_{\Delta}=\{(0,0,0)\}$.
The construction obviously gives us  $\CP^2$ as defined earlier.

\item The (singular) conifold $\calC_0$ is a 3-dimensional affine variety. It is not difficult
to realize that a toric affine variety can only correspond to a single top-dimensional cone in the fan
(see below). The fan for the conifold contains 10 cones (including the 0-dimensional one).
It is generated by four lattice vectors in $N\cong \bbZ^3$:
\be\label{latticeConifold}
v_1= (0,0,1), \quad v_2= (1,0,1), \quad v_3= (1,1,1), \quad v_4= (0,1,1).
\ee
There is a single relation with charge vector $(1,-1,1,-1)$, so $G$ is one dimensional and acts as
\be
G: \, (z_1,z_2,z_3,z_4) \mapsto (\lambda z_1,\lambda^{-1} z_2,\lambda z_3,\lambda^{-1} z_4).
\ee
The zero set is 
\be
Z_{\Delta} = \{z_1=z_3=0\} \cup  \{z_2=z_4=0\}.
\ee
\end{itemize}

\subsection{Coordinate rings and dual cones}
The homogeneous coordinates are very useful for many purposes.
However, it is natural to ask how we can describe a toric variety in \textit{local} coordinates: as for manifolds, we would like to be able to
cover our varieties with open sets equipped with local coordinates. The relevant notion of open sets is different here from the  usual topology of differential geometry\footnote{The natural topology in algebraic geometry is called the Zariski topology. See any textbook such as \cite{Griffiths}.}, but this will not concern us here. We should say, however, that because we deal with singular spaces, the most ``local'' one can get is to affine varieties themselves. This is why it was so crucial to spend some time introducing them. Moreover, because the only non-singular affine variety is $\bbC^m$ itself, for non-singular varieties the relevant open sets are simply $\bbC^m$ and we recover the usual notions for complex manifolds, which we used in section \ref{section CY condutin}.

How do we find such local coordinates?
The fan again provides the answer. To each top-dimensional cone we associate an affine variety (affine patch).
The transition functions between these patches are also naturally encoded in the fan.

Given a single $m$-dimensional cone $\sigma$ spanned by $n$ vectors, we want to find the coordinate ring associated to it.
Since a toric variety is defined as a quotient by $G$, local coordinates should be $G$-invariant polynomials%
\footnote{The reader should generalize the following considerations to the case when $G$ has a non-trivial discrete subgroup $\Gamma$. See the examples below.}:
\be\label{defineGIP}
x= z_1^{n_1}\cdots z_n^{n_n}, \qquad G: \, x \, \mapsto\quad \lambda^{\sum_i Q_i^a n_i} x =x,
\ee
which means that the positive integers $n_i$ are such that $\sum_i Q_i^a n_i=0$.
Because of (\ref{QinkerV}), this means that we can take
\be
n_i = \langle w, v_i \rangle
\ee
for any $w \in \mathrm{Hom}(N,\bbZ)$ : The local coordinates are in one-to-one correspondence with elements in
 \textit{the dual lattice} $M\cong \bbZ^m$,
\be
M = \mathrm{Hom}(N,\bbZ).
\ee
In fact, the condition $n_i\geq 0$ defines the dual real cone $\sigma^\vee \in M_{ \bbR}$,
\be
\sigma^{\vee} = \{ aw  \in M_{\bbR} \quad | \quad a\in \bbR_{\geq 0}, \, \langle w, v_i \rangle \geq 0 \quad \forall v_i \in \sigma  \}.
\ee

Then, the coordinate ring we are looking for is simply 
\be
\calA_{\sigma} \,=\, \bbC[\sigma^{\vee}\cap M].
\ee
Indeed $\sigma^{\vee}\cap M$ is a semi-group defining the monomials in the ring, and the addition in $\sigma^{\vee}\cap M$ becomes
the multiplication in the ring. One can easily write this as the quotient of a polynomial ring by some ideals: 
\begin{itemize}
\item First, find a \textit{minimal} set of lattice vectors $(w_1,\cdots, w_r)$ 
generating $\sigma^{\vee}\cap M$; in general this is the most tricky part of 
the construction. We associate to this set the polynomial ring $\calR_r= \bbC[x_1,\cdots,x_r]$.
\item Find all the relations between the $w_i's$, and associate to each relation an element of $\calR_r$:
\be
\sum_{i\in I} m_i w_i = \sum_{j\in J} m_j w_j   ,\quad m_i, m_j \in\bbN \quad \Rightarrow \quad  p(x)=\prod_{i\in I} x_i^{m_i}- \prod_{i\in J} x_i^{m_i}
\ee
where $I\cup J= \{1,\cdots, r\}$ and $I\cap J=0$. This generates a prime ideal $P_{\sigma}=(p)$, and we then have
\be
\calA_{\sigma} = \bbC[\sigma^{\vee}\cap M] = \frac{\bbC[x_1,\cdots,x_r]}{(p)} \, .
\ee
\end{itemize}
It is not obvious but nonetheless true that this ideal is prime, and moreover it is such that
the associated affine variety
\be
U_{\sigma} = \mathrm{Spec}(\frac{\bbC[x_1,\cdots,x_r]}{P_{\sigma}})
\ee
has dimension $m$%
\footnote{This means that the \textit{height} of the ideal $P_{\sigma}$ is always $r-m$.}. Here we used the notation of (\ref{def of Spectrum}).

The affine varieties $U_{\sigma_i}$, $\sigma_i\in \Delta$, can be patched together to form a more general toric variety $X_{\Delta}$.
Suppose the cone $\tau$ is a \textit{face} of both $\sigma_i$ and $\sigma_j$. Then (exercice),
we have that
\be
\sigma^{\vee}_{i,\, j}\subset \tau^{\vee} \quad\Rightarrow \quad
 \bbC[\sigma_{i,\, j}^{\vee}\cap M] \subset \bbC[\tau^{\vee}\cap M] \quad \Rightarrow \quad U_{\tau}\, \subset\, U_{\sigma_i}\cap U_{\sigma_j}\, .
\ee
In words, the affine set associated to the face is in the intersection of the affine sets of the two cones. Hence the relations between
local coordinates in $x^{(i)}$ for $U_{\sigma_i}$ and $x^{(j)}$ for $U_{\sigma_j}$ can be read off from the relations between the generators
of $\sigma_{i}^{\vee}\cap M$ and $\sigma_{j}^{\vee}\cap M$:
\be\label{transition fcts}
\sum_{l=1}^{r_i} q_l w_l^{(i)} =  \sum_{l'=1}^{r_j} q_{l'} w_{l'}^{(j)}, \quad q_{l,l'}\in \bbZ \quad \Rightarrow \quad
\prod_l^{r_i}(x_l^{(i)})^{q_l} = \prod_{l'}^{r_j} (x_{l'}^{(j)})^{q_{l'}}
\ee
We see that the transition functions are always \textit{rational functions}.
\newline

\begin{figure}[t]
\begin{center}
\subfigure[\small The toric cone for $\bbC^2/\bbZ_2$.]{
\includegraphics[height=5cm]{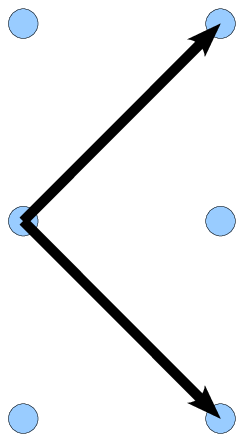}
\label{fanC2Z2}
} \qquad \quad
\subfigure[\small The toric fan for $dP_1$. ]{
 \includegraphics[height=5cm]{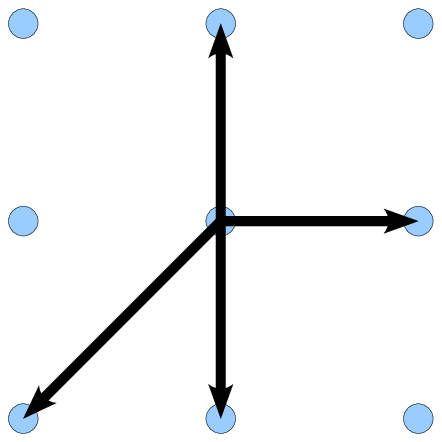}
\label{fandP1}}\quad \qquad
\subfigure[\small The toric fan for $dP_2$. ]{
 \includegraphics[height=5cm]{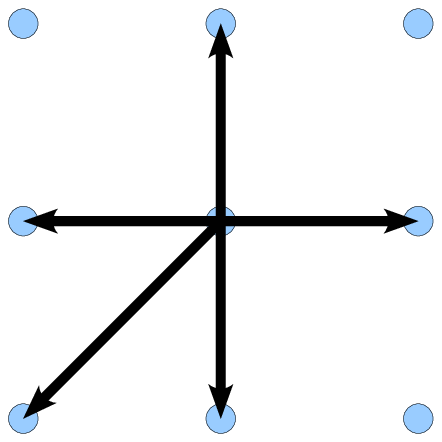}
\label{fandP2}}
\caption{Some examples of toric fans in dimension two.}
\end{center}
\end{figure}

\textbf{Examples:}
\begin{itemize}
\item Consider again the fan for $\CP^2$. There are three 2-dimensional cones, $\sigma_1$, $\sigma_2$, $\sigma_3$, and for each of them
\be
U_{\sigma_i}=\mathrm{Spec}(\bbC[\sigma_i^{\vee}\cap M]) \cong \bbC^2.
\ee
Applying (\ref{transition fcts}), we see that the transition functions between $U_{\sigma_1}=(x_1,x_2)$ and $U_{\sigma_2}=(y_1,y_2)$, for instance, are
\be
x_1= \frac{y_1}{y_2}, \qquad x_2= \frac{1}{y_2}.
\ee

\item Consider the simple fan in $N\cong \bbZ_2$ shown in the Fig.\ref{fanC2Z2}. It has a single top dimensional cone, spanned by
\be
v_1= (1,1), \qquad \mathrm{et}\quad v_2=(1,-1).
\ee
Notice that there is no relation between the two vectors, so $\tilde{G}$ is trivial, however we do have a discrete group $\Gamma$ in the quotient (\ref{ConstruTV1}), $\Gamma= \bbZ_2$, since $v_1$ and $v_2$ only generates half of the lattice $N$. In term of local coordinates, we have the dual cone $\sigma^{\vee}$ generated by $w_1= (1,-1)$ and $w_2= (1,1)$. In order to generate the dual cone $\sigma^{\vee}\cap M$ (over $\bbZ$), we need to introduce a third vector $w_3=(1,0)$. Then, assigning homogeneous coordinates $x$,$y$,$z$ to these three vectors, we have the relation
\be
w_1 +w_2=2w_3 \qquad \Leftrightarrow \qquad xy=z^2\, .
\ee
The later equation is the algebraic definition of $\bbC^2/\bbZ_2$, seen as an affine variety.

\textbf{Exercice:} Make sure you can check this last claim. The $\bbZ_2$ group acts on $\bbC^2$ as $(z_1,z_2)\rightarrow(-z_1,-z_2)$. You just have to build invariants under the orbifold action, and check they indeed correspond to the above local coordinates.

\item In Figures \ref{fandP1} and \ref{fandP2}, we have drawn the toric fans for the first and second del Pezzo surfaces (denoted $dP_1$ and $dP_2$). As you can see from the fan, they are smooth surfaces (each dual cone corresponds to a $\bbC^2$ patch). You should be able to work out the transition functions between the patches as in the case of $\CP^2$. 

\item Exercice: Find the local coordinates for the conifold $\calC_0$ using the procedure of this subsection. You should find the affine variety (\ref{focofCOniiii}).
\end{itemize}

Now comes an important proposition:\newline

\textbf{Proposition: } A toric variety $X_{\Delta}$ is \textit{compact} if and only if its fan $\Delta$ spans the whole $N_{\bbR}$.\newline

See Chapter 2 of \cite{Fulton} for a proof. One sees in the above examples that $\CP^2$ and $dP_{1,2}$ are compact spaces, while $\bbC^2/\bbZ_2$ or the conifold are of course  not.

\subsection{Calabi-Yau toric varieties}
In this subsection, we show how the Calabi-Yau condition is translated into a simple condition on
 the combinatoric data for $X_{\Delta}$.

We saw in section \ref{section CY condutin} that the Calabi-Yau condition for $X$ is the triviality of the canonical bundle $K_X$.
Here we show how one can express $K_X$ in term of a simple set of divisors called toric divisors.\newline

\textbf{Definition: } A \textit{toric divisor} is a divisor invariant under the action of $G$.\newline

Using the homogeneous coordinates $(z_i)$, we can easily define subvarieties that are G-invariant. Indeed, 
the simple algebraic sets
\be
\{(z_1,\cdots,z_n)\quad | \quad  z_i=0 \, \forall i\in I\subset \{1,\cdots,n\} \}.
\ee
are obviously G-invariant. In particular, the subvarieties
\be
D_i \equiv \{z_i =0\}\cap X_{\Delta}
\ee
are toric divisors \footnote{This is because the ideal $(z_i)$ has height one, which implies $D_i$ is codimension one in $X_{\Delta}$ too.}.
They actually generate the full group of divisors of $X_{\Delta}$.

Consider $X_{\Delta}$ \textit{smooth} with canonical bundle $K_X$. One can show that
\be\label{relKxtoDivD}
K_X = \calO_X(-\sum_i^n D_i)\, .
\ee
The argument goes as follows. Because $X_{\Delta}$ is regular, each coordinate ring $\calA_{\sigma}$ is freely generated:
\be
U_{\sigma} \cong \bbC^k \times (\bbC^*)^{m-k}, \quad \Leftrightarrow \quad \calA_{\sigma} = \bbC[x_1,\cdots, x_k, x_{k+1},x_{k+1}^{-1}, \cdots, x_m,x_m^{-1}]\, .
\ee
Consider for simplicity the case $k=m$, which means $\sigma$ is of dimension $m$ (the generalization is straightforward).
A section of the canonical bundle is
\be
\Omega = \frac{1}{x_1\cdots x_m}\dd x_1\wedge\cdots \wedge \dd x_m.
\ee
This section corresponds to a divisor. Equivalently, the dual
section in $K_X^{-1}$ corresponds to an \textit{effective} divisor, described locally by
\be
\{x_1 \cdots x_m=0\}\cap U_{\sigma}.
\ee
It is called the anti-canonical divisor.

On the other hand, a section of the line bundle $\calO(\sum_i D_i)$ corresponds to the divisor
\be
\{ \{z_1z_2\cdots z_n=0 \}\cap X\} \subset X.
\ee

We know that
\be
x_1\cdots x_m \, = \, z_1^{\langle w, v_1\rangle}\cdots z_n^{\langle w, v_n\rangle} \quad \mathrm{with} \quad w= \sum_j^m w_j.
\ee
Suppose the first $m$ vectors amongst the $v_i$'s span the cone $\sigma$.
Since $U_{\sigma}\cong \bbC^m$, we have $\langle w, v_i\rangle=1$ for $i=1,\cdots,m$. 
Hence the anti-canonical divisor corresponds to $\sum_i D_i$ on $U_{\sigma}$.
This implies that $K^{-1}_X= \calO(\sum_i^n D_i)$, which is what we wanted to show.
\newline

\textbf{Exercices: } 
\begin{itemize} 
\item Work out the relation explicitly for $\CP^2$.
\item Work out the relation between $K_X$ and $\calO(-\sum_i D_i)$ for the \textit{singular} conifold $\calC_0$. Does (\ref{relKxtoDivD}) hold ?
\end{itemize}

The important relation (\ref{relKxtoDivD}) allows us to state the Calabi-Yau condition (triviality of the canonical bundle) 
in a very simple way. Note that any G-invariant function, as defined in (\ref{defineGIP}), is of course a section of the trivial bundle.
We then see that $\calO_X(\sum_i D_i)$ is trivial if and only if
\be
G: z_1\cdots z_n \mapsto\, \lambda^{\sum_i Q^a_i} (z_1\cdots z_n)= z_1\cdots z_n \quad \Leftrightarrow \quad \sum_i Q^a_i=0,
\ee
or equivalently if there exist a dual vector $w\in M$ such that $\langle w, v_i\rangle=1$ for all $v_i$ in the fan. 
We then have shown the following:\newline

\textbf{Proposition:} The toric variety $X_{\Delta}$ is Calabi-Yau if and only if all the vectors $v_i$ in $\Delta$
end on the same hyperplane in $N$, which happens if and only if $\sum_i Q^a_i=0$ $\forall a$.  \newline

Remark that we chose the $v_i$ for the conifold in (\ref{latticeConifold}) especially to make the CY property explicit.

It also follows from the proposition at the end of the last subsection that a toric CY cannot be compact.

\subsection{Toric diagrams and p-q webs}
For toric Calabi-Yau varieties, the combinatoric information encoded in the fan can be expressed in term of a reduced 
lattice of dimension $m-1$.

This is particularly convenient in order to describe toric CY threefolds (toric CY of dimension 3), which are the objects of main relevance
to physics. Instead of drawing a 3-dimensional fan, we can simply project it on the special plane defined by $\langle w, v_i\rangle=1$.
\newline

\begin{figure}[t]
\begin{center}
\subfigure[\small Toric diagram for the conifold.]{
\includegraphics[height=3cm]{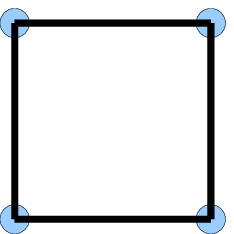}
\label{diagConi}
} \qquad \qquad
\subfigure[\small Toric diagram for $C_{\bbC}(dP_1)$.]{
\includegraphics[height=4cm]{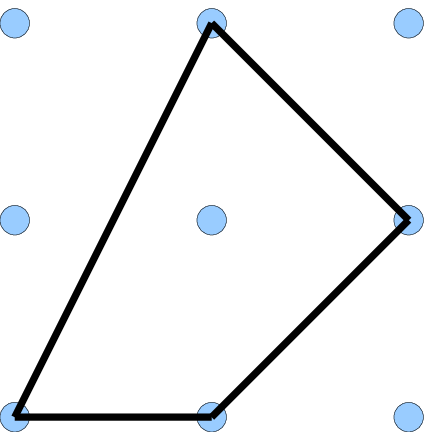}
\label{diagdP1}
} \qquad \qquad
\subfigure[\small Toric diagram for the SPP.]{
\includegraphics[height=4cm]{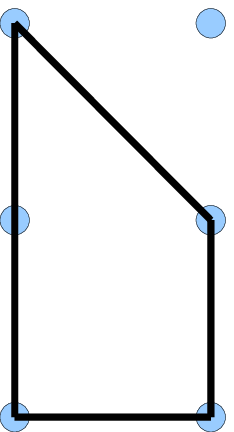}
\label{diagSPP}
} 
\caption{Some examples of toric diagrams for local CY threefolds.}
\end{center}
\end{figure}

\begin{figure}[t]
\begin{center}
\subfigure[\small pq-web of the resolved conifold.]{
\includegraphics[height=4cm]{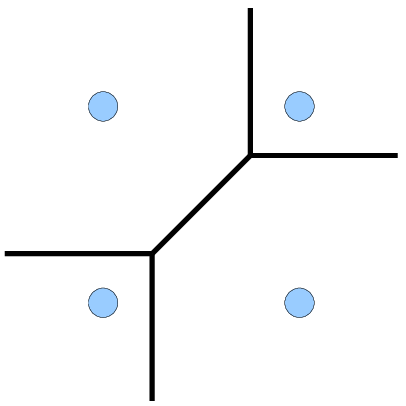}
\label{pqConi}
} \qquad \qquad\qquad
\subfigure[\small pq-web of the (resolved) $C_{\bbC}(dP_1)$.]{
\includegraphics[height=4.5cm]{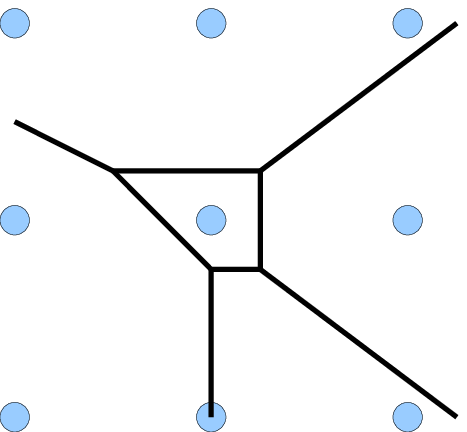}
\label{pqCdP1}
}
\caption{pq-webs}
\end{center}
\end{figure}

In the Figures are some examples of toric diagram. The one for the conifold is given in Fig.\ref{diagConi}, while Fig.\ref{diagdP1} corresponds to the complex cone over the $dP_1$ surface, which happens to be a Calabi-Yau singularity.

In Fig.\ref{diagSPP} is a singularity called the Suspended Pinch Point (SPP). As an exercice, you can work out the toric description of the SPP. For instance, show that in local coordinates, the SPP is an affine variety in $\bbC^4$ defined by the ideal $(xy-z^2t)$ in $\bbC[x,y,z,t]$.

One can also draw the dual of the toric diagram, which is called the pq-web (simply, for each line in the toric diagram, you draw an orthogonal line in the pq-web). Such webs have a nice physical interpretation as webs intersecting fivebranes \cite{Aharony:1997bh}. 

 The examples of the conifold and of the first del Pezzo cone are given in  Figs \ref{pqConi} and \ref{pqCdP1}. You first have to triangulate the diagram (see next Section), and then take the dual diagram.

\section{Dealing with toric singularities}\label{Section: singularities}

We are now ready to deal with singularities in toric geometry. In physics, singularities are usually the signal of a breakdown of our theories: for instance the self-energy of a classical point charge is infinite, but we know of a way to construct a coherent theory of ``point charges'', namely quantum field theory.
Another class of examples, of a tougher kind, are the classical singularities in general relativity. In that case too we expect them to be  artifacts of the classical description, while a consistent theory of quantum gravity would do away with them.

String theory is the best candidate we have for such a quantum theory. There we know of some phenomena of singularity resolution through quantum effects, as for instance in \cite{Strominger:1995cz} which crucially relied on  properties of the conifold geometry.

On the other hand, from a mathematical point of view, one simple way to understand singular spaces is to ``resolve'' or ``deform'' their singularities - we will define both these notions momentarily. One might hope that the ``slightly deformed'' space is similar to the original one: from the point of view of algebraic geometry this is wrong in general, because that kind of singularity resolution process is hardly ever a continuous process. However, it is often the best way we have to understand singularities.

Here we focus on the simple concepts of resolution and deformation of toric Calabi-Yau singularities in algebraic geometry, as these processes are often good toy models of poorly understood string theory phenomena. For instance the deformation of the conifold played a crucial role in the extension of the $AdS$/CFT correspondence to more general setups \cite{Klebanov:2000hb}.\newline

What is a singularity in algebraic geometry? Let $X$ be an algebraic variety of dimension $m$. A point in $X$ will be 
deemed singular if the tangent space at that point has dimension \textit{larger} than $\mathrm{dim}X=m$.

Without loss of generality, we can define the tangent space $T_x X$ at the point $x$
for affine varieties only:
\paragraph{Tangent space of $X$.} If $X=Z(\calJ)$, with $\calJ$ a prime ideal of $\calR_n= \bbC[x_1,\cdots,x_n]$, 
we can define the following ideal of $\calR_n$, generated by degree one polynomials, for each point $x$:
\be
\calJ_x =\big\{\sum_i^n \frac{\partial p}{\partial x_i}(x)\, \big(x_i-x_i(x)\big) \, \in \calR_n  \,|\,  p\in \calJ \big\}.
\ee
This ideal generates a \textit{linear} affine variety that we define to be the tangent space at $x\in X$,
\be
T_xX \equiv Z(\calJ_x).
\ee

This obviously generalizes the usual definition of a tangent space. 
Now, a point $x$ in $X$ is called non-singular if its tangent space has the same dimension as the variety $X$.
Of course, $X$ is said to be non-singular if it has no singular points. For singular points the dimension of $T_xX$ is larger than $m$.

\textbf{Exercice:} Compute the tangent space of the conifold $x_1x_2-x_3x_4=0$, both at the singularity $x_i=0$, and away from it. 
\newline

Practically speaking, when given an affine variety in terms of its defining polynomials $p(x)$ (i.e. in local coordinates), one finds the \textit{singular locus} as the set of points $x$ such that
\be
p(x)=0, \qquad dp(x)=0\, .
\ee

For toric varieties, there is a straightforward \textbf{theorem}  \cite{Fulton} which states that the affine variety $X_{\sigma}$ associated to the cone $\sigma$ is non-singular if and only if $\sigma$ is generated by an integral basis of the lattice $N$.

\paragraph{Polytope and unit simplex.}  In $m$ dimensions, we will call polytope  the convex hull\footnote{As one can find in Wikipedia.org, for instance, a convex hull of $k$ points is the minimal convex set containing these points. This is just the higher dimensional generalization of 2-dimensional polygons and 3-dimensional polyhedrons.} of $k$ distinct points in $N\cong \bbZ^m$.
Given a $m$-dimensional cone $\sigma$ in a toric fan, the basic polytope is the polytope delimited by the origin and the vectors $v_i \in \sigma$. For instance, for the conifold we have $(0,0,0)$, $(0,0,1)$, $(1,0,1)$, $(1,1,1)$, $(0,1,1)$. In general we have $n$ vectors $v_i$, so we have $k=n+1$ points defining the basic polytope.

On the other hand, a simplex is the $m$-dimensional generalization of a triangle or tetrahedron: the convex hull of $m+1$ points. We define the simplicial volume of a polytope as the number of simplexes it contains.

Indeed, any polytope can be subdivided into simplexes: this is called a simplicial decomposition. 
We can now reformulate the above theorem (exercice) as:

\paragraph{Proposition:} The affine variety $X_{\sigma}$ associated to the cone $\sigma$ is non-singular if and only if the basic polytope associated to $\sigma\cap N$ has unit simplicial volume.

\subsection{Resolution of toric singularities and simplicial decomposition}
We can then ``desingularize'' any toric variety by subdividing its associated fan further until every cone is based on a
unit simplex.\newline

\begin{figure}[t]
\begin{center}
\subfigure{
\includegraphics[height=3cm]{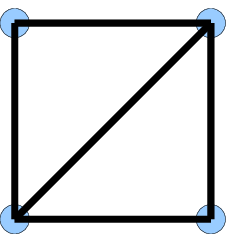}
\label{resConi1}
} \qquad \qquad\qquad
\subfigure{
\includegraphics[height=3cm]{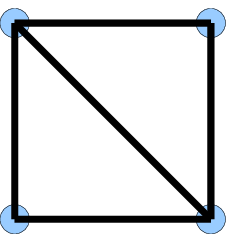}
\label{resConi2}
}
\caption{The two possible resolutions of the conifold.\label{Resolu of Con}}
\end{center}
\end{figure}

For a toric CY threefold, simplicial decomposition is equivalent to a \textit{triangulation} of its toric diagram. For instance, in Fig.\ref{diagConi} one can see that the basic simplex of $\calC_0$ has simplicial volume 2, while the cone over $dP_1$ in Fig.\ref{diagdP1} has simplicial volume 4, so they are singular. The two possible triangulations of the conifold diagram are shown in Figure \ref{Resolu of Con}.

\textbf{Example.} Take the conifold again. Its basic simplex has volume $2$. We can split it into
a fan containing two cones, each of unit volume. This is called the \textit{resolved} conifold. 
Now we have two 3-dimensional cones in the fan, $\sigma_1$ and $\sigma_2$. The dual cones correspond
to two copies of $\bbC^3$:
\bea
\sigma^{\vee}_1 \quad : \quad \{(1,0,0),\, (0,-1,1),\, (-1,1,0)\} \qquad &\rightarrow & \bbC^3 =\{x_1,y_1,z_1\},\\
\sigma^{\vee}_2 \quad : \quad \{(0,1,0),\, (-1,0,1),\, (1,-1,0)\} \qquad &\rightarrow & \bbC^3 =\{x_2,y_2,z_2\},
\eea
We see that the relations between the vectors in the dual lattice give us the following transition functions
between the two patches:
\be
z_1=\frac{1}{z_2}, \qquad  \frac{x_1}{y_1}=\frac{x_2}{y_2}, \qquad x_1 z_1=x_2, \quad etc.
\ee
The second relation is actually the defining equation of the conifold singularity. Before the triangulation
of the toric diagram, that was all what one would get. The triangulation procedure introduced new coordinates,
$z_1$ and $z_2$ with $z_1=1/z_2$, which give the complex
structure of a $\CP^1$. Away from the point $x_1=y_1=x_2=y_2=0$, these coordinates are redundant, but at the former conifold singularity,
we now have a full $\CP^1$.

Remark that in the homogeneous coordinate description, you still have the same four coordinates $z_1,\cdots z_4$. What changes is that the zero set $Z_{\Delta}$ is now different when the fan is subdivided: $Z_{\Delta}=\{z_1=z_2=z_3=z_4=0\}$, so that the singularity is effectively removed. 
\newline

Such a procedure, which replaces an isolated singularity by a holomorphic cycle, is called a \textit{resolution} of the singularity.\newline

More precisely \cite{joyce}, a resolution $(\tilde{X},\pi)$ of the variety $X$ is a non-singular variety  $\tilde{X}$ together with a surjective map $\pi: \tilde{X}\rightarrow X$ which is biholomorphic on open sets wherever $\pi$ is also injective. In other words, $\pi$ is a biholomorphism ``away'' from the singular points,
while the singularities are replaced by some smooth spaces, for instance by means of a small resolution, or by blowing them up.

\paragraph{Blow up.} 
A blow up is a procedure which replaces the singular locus of $X$ by $\CP^{m-1}$. (Beware that in the physics literature the terms ``blow up'' is sometimes used to designate any kind of resolution.) Hence a blow up introduces new divisors, called \textit{exceptional divisors} (these are defined as the prime divisors $E\subset \tilde{X}$ such that $\pi (E)$ has codimension 2 or more in $X$). 

\paragraph{Small resolution.}
On the other hand, a small resolution is  a resolution such that $\tilde{X}$ has 
no exceptional divisors. In particular, the resolution of the conifold is a small resolution.\newline

The resolutions we usually deal with in string theory are actually \textit{crepant resolutions}. The resolution $(\tilde{X}, \pi)$ of X
is said to be crepant when%
\footnote{The canonical bundle for 
a singular variety is itself tricky to define. A straightforward generalisation of the idea of holomorphic line bundle is 
what is called an \textit{invertible sheaf} (which is a sheaf of modules locally isomorphic to the structure sheaf $\calO_X$).
Then one works with the sheaf $\calO(K_X)$, the sheaf of regular sections of $K_X$, which is assumed to be invertible.
You can pull-back this sheaf using $\pi$, but in general $\pi^*(\calO(K_X))$ is not equal to $\calO(K_{\tilde{X}})$. 
It turns out that the discrepancy can come from exceptional 
divisors only, and if there is no discrepancy the resolution is said to be \textit{crepant}
(so we see that small resolutions are crepant by definition).}
\be
\pi^*(\calO(K_X)) = \calO(K_{\tilde{X}}).
\ee
In particular, the Calabi-Yau condition is preserved by a crepant resolution.\newline

 For a toric CY threefold, a blow up consists in introducing a $\CP^2$ 
at the singularity, while a small resolution introduces a $\CP^1$ instead. You can convince
yourself (exercice) that the blow up corresponds to adding an internal point in the toric diagram (see the pq-web Fig.\ref{pqCdP1} for instance),
while the small resolution corresponds to a triangulation which does not introduce new points (like for the conifold).

\subsection{Deformation of toric singularities: the versal space }
Another way to get rid of a singularity is to \textit{deform} it: this modifies the complex structure.
For instance, we saw that the conifold equation $x_1x_2-x_3x_4=0$ admits a deformation to
\be\label{def of Coni exp}
x_1x_2-x_3x_4= e, \qquad e\neq 0.
\ee
This new space, called the \textit{deformed conifold}, is non-singular. The complex structure has
obviously changed, but it turns out that it is still a Calabi-Yau variety. However, it is \textit{not}
toric anymore, because the deformation has broken one of the $\bbC^*$ action in the $\bbT^3$ acting on the 
singular conifold (as one can see from the equation).
In this particular case, the Calabi-Yau metric is explicitly known \cite{Candelas:1989js}
. 

\begin{figure}[t]
\begin{center}
\includegraphics[height=3cm]{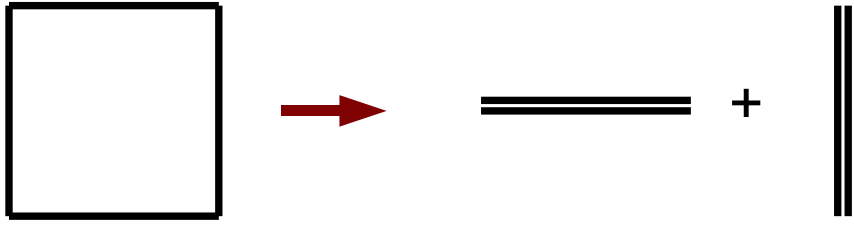}
\caption{Splitting of the conifold diagram into Minkowski summands.\label{defConDiag}}
\end{center}
\end{figure}

It turns out
that for any deformation of the
defining polynomials which is of degree lower or equal to these same polynomials, the resulting
deformed variety is still Calabi-Yau. Of course in general we don't know the Ricci-flat metric on it,
but the CY theorem guarantees its existence.

Since we are dealing here with non-compact CY varieties, we also should not modify the boundary conditions at
infinity. This means that we focus on \textit{normalizable deformations}, which are those which do not change the 
defining polynomials at infinity. 

\paragraph{Exercice.} Consider the following variety (called the ``suspended pinch point'', or SPP),
\be
x_1x_2 -x_3(x_4)^2=0.
\ee
Show that its singular locus is the line
\be
x_1=x_2=x_4=0, \forall x_3.
\ee
Find all the deformations (and check if it is normalizable)
\newline

For a single intersection variety like the one above, it is easy to work out by hand all the possible deformations.
For more complicated varieties, however, it becomes tedious. Also, for non-complete intersection varieties
\footnote{ One talks of non-complete intersection when the dimension of the embedding space $\bbC^n$ minus
 the number of defining polynomials is smaller than the dimension of $X$. It is the general case. (In algebraic language,
 it means that the height of the defining ideal is smaller than the number of generating polynomials.)},
  it may happen that there is no consistent modification of the defining equations. 

\begin{figure}[t]
\begin{center}
\includegraphics[height=3cm]{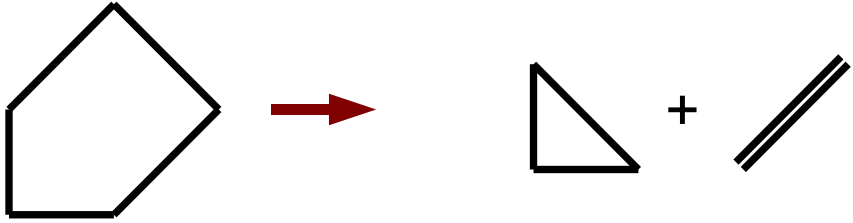}
\caption{Splitting of the $C_{\bbC}(dP_2)$ diagram into Minkowsky summands. The triangle corresponds to a remaining singularity which admits no complex deformation. \label{defdP2Diag}}
\end{center}
\end{figure}

For toric varieties, there exists a very useful algorithm, due to Altmann \cite{Altmann1}, which
gives the number of  normalizable deformations of the singularity for any \textit{isolated toric CY singularity} (and also their explicit form, see \cite{Altmann1}, or 
\cite{Pinansky,Argurio:2007vq} for some physics papers which use it in detail).%
\footnote{Notice that the SPP in the example above is not an isolated singularities: it has a full $\bbC$ worth of singularities, a singularity line.}
 We will focus on CY threefolds,
that we can draw as toric diagrams on a sheet of paper, and where all the interesting phenomenons occur.

The various complex deformations of an isolated CY singularity correspond to the possible ``Minkowski decompositions''
of the toric diagram. This means that we deform the toric diagram into closed sub-diagrams (called Minkowski summands). See Figures \ref{defConDiag} and \ref{defdP2Diag}.
What we are really looking for are the ``breathing modes'' of the toric diagram. We do it in the following way:
\begin{itemize}
\item Consider an affine toric Calabi-Yau threefold, with its toric diagram $D$ containing $n$ points and $n$ edges.
First, assign to each edge of $D\subset \bbZ^2$ a lattice vector 
\be
d^i = p_h- p_t,
\ee
given by the difference between the head and the tail of the corresponding edge of $D$, when going 
in the counterclockwise direction. 
\item Define the vector space
\be\label{lin space of def}
V(D) = \big\{(t_1,\cdots, t_{n}) \, | \, \sum_i^{n} t_i d^i=0 \big\}\, .
\ee
This vector space, including the trivial $(t,t,t,t)$ component, is obviously of dimension $n-2$. Ignoring the trivial rescaling, this is the linearized space of deformations of $X$, of dimension $n-3$.
The deformation could be obstructed at second order, however.
\item The \textit{versal \footnote{``Versal space'' means that all the possible deformations are there, but that the same deformation
might appear several times (if it appear only once we would have a ``universal'' space of deformation, that is what happens
for compact Calabi-Yau varieties, whose complex moduli space has a simpler topology. See \cite{joyce}.).}
 space of complex deformations of $X$} is defined by the following ideal of $\bbC [t_1,\cdots,t_{n}]$:
\be
\calJ = \Big(p_k \equiv\sum_i^{n} (t_i)^k d^i\, |\, k\in \bbZ_{>0}\Big).
\ee
Actually this ideal is generated by the finite set of polynomials  $p_1, \cdots , p_K$, where $K$ is the maximum of the lattice 
width of the minimal pair of strips containing $D$ \cite{Altmann1}.
\end{itemize}

This whole procedure amounts to find the Minkowski summands of the diagram $D$. In term of the dual pq-web, it corresponds
to splitting the pq-web into sub-webs in equilibrium (i.e. the external legs must still sum to zero). For instance, you can see that the diagram in Fig.\ref{diagdP1} admits no Minkowski decomposition. This means that the $dP_1$ singularity cannot be deformed: although its linear space of deformations (\ref{lin space of def}) has dimension one, there is an obstruction at second order.

\paragraph{Example.} Consider the conifold, whose diagram is just a square. We have the following edge vectors:
\be
d^1= (1,0), \quad d^2=(0,1), \quad d^3= (-1,0), \quad d^4 = (0,-1)\, .
\ee
The linear space of deformation is simply generated by $(t_1,t_2,t_1,t_2)$. There is no higher order obstruction so the versal space boils down to the linear space
\be
\mathrm{Spec}(\bbC[t]) = \bbC\, ,
\ee
corresponding to the freedom of adding a constant term $e$ in (\ref{def of Coni exp}).

\paragraph{Exercice.} Work out the versal space of deformations for the complex cone over $dP_2$. (See Section 4 of \cite{Pinansky} and Appendix B of \cite{Argurio:2007vq} for more details.)

\section{Toric geometry 2: Gauged linear sigma-model}\label{section GLSM}

There is an alternative, complementary approach to toric varieties, which does not directly rely on algebraic geometry,
but rather deals with the \textit{symplectic} or (more precisely) \textit{K\"ahler} properties of our varieties.

The idea is to split the quotient by $(\bbC^*)^{n-m}$ in (\ref{ConstruTV1}) into two steps. 
Since 
\be
\bbC^* \cong U(1) \times \bbR_{>0},
\ee
we will first fix some ``point'' $t \in \bbR_{>0}$ (and $t \rightarrow 0$ will correspond to a singular limit for the toric variety),
and secondly we will divide by the $U(1)$ action (which is the gauge group, in the physics parlance).
Such a procedure is well defined because we have a well defined K\"ahler form on the parent space $\bbC^n$. It is called a 
\textit{K\"ahler quotient} of $\bbC^n$.

\subsection{K\"ahler quotient and moment maps}
Before exploring  the ``physics'', let us briefly explain what is a K\"ahler quotient mathematically. We will focus on the quotient of $\bbC^n$ by an abelian group.
The group $G=U(1)^r$ ($r=n-m$) acts on $\bbC^n$ as  (compare to (\ref{action of Gtilde}))
\be\label{action of U1 gauge group}
U(1)^r : \bbC^n\rightarrow \bbC^n\quad  :\quad  (z_i) \mapsto (e^{i\sum_a\xi_a Q^a_i} z_i),
\ee
where $\xi = (\xi_1,\cdots,\xi_r)$ are element of the Lie algebra $g=u(1)^r \cong \bbR^r$ of the gauge group.
The action of $g$ is then
\be\label{vectXI}
\xi \cdot F(z) = \big(i\sum_{i}\sum_a \xi_a Q^a_i z_i \frac{\partial}{\partial z_i}+ c.c.)F(z).
\ee
The complex conjugate is necessary to make it a real action.

\paragraph{Definition: } Given a K\"ahler manifold with K\"ahler 2-form $\omega$,
a \textit{moment map} $\mu$ for the group action of $G$ on $\bbC^n$ is an element
of the dual Lie algebra, $g^*$, such that
\be
d \langle \mu, \xi \rangle \equiv d (\mu^a\xi_a) = i_\xi \omega,
\ee
where here $i_{\xi}$ denote the interior product with the vector $\xi$ appearing on the r.h.s.
of (\ref{vectXI}).

\paragraph{Exercice.} Guess why it is called a moment map. 
(Hint: the function $\mu^a\xi_a$ can be seen as a Hamiltonian on a symplectic 
manifold describing a phase-space in classical mechanics.) You could also check that
the existence of a moment map implies that the $G$-action preserves $\omega$ as well as
the complex structure (the Lie derivative of both w.r.t. $\xi$ being zero), so that the elements of $g$ really correspond to
holomorphic Killing vectors.\newline

You can easily show that, in our case, the  K\"ahler manifold being simply $\bbC^n$ with the canonical K\"ahler form
\be
\omega = -i \sum_i dz_i\wedge d\bar{z}_i\, ,
\ee
 the moment maps are
\be
\mu^a = \sum_i Q^a_i |z_i|^2 -t^a, 
\ee
where the $t^a$ are integration constants. Then, the K\"ahler quotient proceeds as follows:
\begin{itemize}
\item Set $\mu^a =0$, i.e. 
\be\label{DtermsGLSM}
\sum_i Q^a_i |z_i|^2  = t^a \qquad \forall a,
\ee
This is called a restriction to a level set at level $t$. The parameters $t^a$ could be set to zero, as we will see.

\item The second step is to quotient by the compact gauge group $U(1)^r$, whose action was defined in (\ref{action of U1 gauge group}).
\end{itemize}

The first step defines a lower-dimensional \textit{real} algebraic submanifold in the space  $\bbR^n_{\geq 0}$
spanned by the $|z_i|$'s. Then the second step tells us which subgroup $U(1)^m$ of the $U(1)^n \subset \bbC^n$
torus must be fibered at each point to produce the final $m$-dimensional variety.

\subsection{The GLSM story}
If your are familiar with supersymmetric theories (or you have read the introduction carefully), the above must have looked like known territory. 
The restriction to a level set is simply the imposition of the D-term constraints in some abelian gauge theory, while the second steps
corresponds to fixing the gauge freedom (restricting to gauge orbits).

Hence, we can see toric varieties as the \textit{moduli space of vacua} of a ``gauged linear sigma-model'' (GLSM). 
We have $n$ chiral fields whose scalar component are the $z_i$'s, and they are charged under the gauge group as
\be\nonumber
\begin{tabular}{l|ccc|c}
 & $z_1$ &$\cdots$ & $z_n$ & FI \\
 \hline
  $U(1)_1$  &$Q^1_1$ &$\cdots$ & $Q^1_n$ & $t^1$\\
    $\vdots$  & &$\ddots$ && \vdots\\
      $U(1)_r$  &$Q^r_1$ &$\cdots$ & $Q^r_n$&  $t^r$
\end{tabular}
\ee
Because the gauge group is $U(1)^r$, there are possible Fayet-Iliopoulos (FI) parameters $t^a$ in the D-terms conditions 
(\ref{DtermsGLSM}). 
This was first realised by Witten in \cite{Witten:1993yc}, where he used a 2-dimensional GLSM as an auxiliary device 
to find 2-dimensional CFTs.
Here the auxiliary theory is four dimensional (the main difference with respect to \cite{Witten:1993yc} being that the FI
parameters are real), and its infrared corresponds to a Calabi-Yau ``as probed by D3-branes''\footnote{You should not take this analogy too seriously: the GLSM is an auxiliary construction, like the fan, there is \textit{a priori} no real physics there.}.

\paragraph{Examples.} Consider the GLSM with a single $U(1)$ and four fields with the following charges:
\be\nonumber
\begin{tabular}{l|cccc}
 & $z_1$ &$z_2$ & $z_3$ & $z_4$ \\
 \hline
  $U(1)$  &$p$ &$p$ & $-p+q$ & $-p-q$ 
\end{tabular}
\ee
and no FI term. The resulting toric Calabi-Yau singularity is a real cone over a real 5-dimensional Sasaki-Einstein\footnote{Sasaki basically means that the real 6-dimensional cone is K\"ahler, while the Einstein condition on the 5-dimensional base metric implies the Ricci-flatness of the cone. Hence a SE manifold of real dimension $2n-1$ is the real base of a CY cone of complex dimension $n$.}
 space called $Y^{p,q}$ \cite{Martelli:2004wu}. This family of toric CY singularities has received a lot a attention in the physics litterature during the last years, because the corresponding Ricci-flat metrics are known explicitly \cite{Gauntlett:2004yd}, which is a rather spectacular feat and allowed for some new checks of the AdS/CFT correspondence.

Exercice: Check that the complex cone over $dP_1$ discussed before is actually the real cone over $Y^{2,1}$. It suffices to find the GLSM charges from the toric diagram in Fig.\ref{diagdP1}.

\subsection{Toric varieties as torus fibration of polytopes}
An affine toric variety $X$ can be visualized quite simply as a torus fibration over a \textit{polytope} $\Delta$:
\be\label{exse1111}
U(1)^m \quad\rightarrow\quad X \quad \stackrel{\mu_H}{\longrightarrow} \quad \Delta
\ee
Indeed, the toric variety $X$ has an isometry group 
\be
H= U(1)^m = \frac{U(1)^n}{U(1)^r},
\ee
and there is a moment map $\mu_H$ on $X$ associated to this $H$. This moment map is
precisely the map which projects out the $U(1)^m$ fibers in the exact short sequence (\ref{exse1111}) \cite{Martelli:2004wu}. 

For cones ($t^a=0$), the polytope is precisely the toric cone $\sigma$ for $X$.
Given the charge vectors $Q^a$, one can construct several $\sigma$'s such that
\be
\sigma = \big\{ \mathrm{Span}((v_i) \in N\cong \bbZ^m) \, | \, \sum_i Q_i^a v_i \, \forall a   \big\}.
\ee
All these $\sigma$ are related to each other by $Sl(m,\bbZ)$ transformations.
Consider, for instance, taking an orthogonal basis of $\bbZ^m$ for the first $m$ lattice vectors
$v_i$ (corresponding to the first $m$ homogeneous coordinates). The remaining vectors
of $\sigma$ follow from (\ref{QinkerV}). This choice of lattice basis vectors
corresponds to a choice of subgroup for 
\be
U(1)^m \subset U(1)^n\, .
\ee 
This is simply because we made a choice about which of the homogeneous coordinates $(z_i)$ are the ``dependent'' ones.
Here we chose the variables $(z_{m+1}, \cdots, z_{n})$ to be functions of the $(z_1, \cdots, z_m)$.
More precisely, the modulus are fixed by the D-terms (\ref{DtermsGLSM}), while the phases of $(z_{m+1}, \cdots, z_{n})$
are redundant $U(1)^r$ degrees of freedom that we can gauge fix.

Then we see explicitly that the affine toric  variety is realized as a $U(1)^m$ fibration of $\sigma \subset N_{\bbR}$.
In the bulk, the real torus $T^m\cong U(1)^m$ is non-degenerate, while on the intersection of the
hyperplane $(v_j= 0,\, \forall j\in I \subset (1,\cdots m))$ with the cone $\sigma$,
there is a degeneration of the $(\#I)$-torus $(T^1_{j_1}\times \cdots \times T^1_{(\#I)})$.
At the tip of $\sigma$ the whole torus shrinks to zero, and we have a singularity.

\paragraph{Example.} Take the conifold again. We have $Q=(1,-1,1,-1)$. If we take a basis of 
$\bbZ^3$ as $(0,0,1)$, $(0,1,1)$ and $(1,1,1)$, we must have that the fourth vector in $\sigma$
be $(1,0,1)$. On the other hand, if we take the orthogonal basis for the lattice, $\sigma$ is 
generated by
\be
(1,0,0),\quad (0,1,0),\quad (0,0,1),\quad (1,-1,1).
\ee
We see that the first $\sigma$ is obtained from the second by the $Sl(3,\bbZ)$ transformation
\be
\left(\begin{array}{ccc} 0&1 &1 \\0&0&1\\1&1&1 \end{array}\right).
\ee
\textbf{Exercise:} work out the different ways the torus $T^3$ degenerates on the boundaries of $\sigma$.\newline

When $t^a\neq 0$ for some $a$'s, we have a (possibly partially) \textit{resolved} singularity.
From the polytope point of view, the resolution amounts to ``chopping off'' the tip of $\sigma$,
since we cannot reach the point $z_1=\cdots =z_n=0$ anymore
(Exercise: visualize this for the conifold.) 
As an aside, let us note the interesting relation between the parameter $t^a$
and the period of the K\"ahler form on the corresponding 2-cycle $C^a$
(in the case of a small resolution by a $\CP^1$) \cite{Denef:2008wq} :
\be
\int_{C^a} \omega = t^a.
\ee
So the FI parameters in the GLSM really map to the ``K\"ahler volumes'' of the resolving cycles.


The GLSM perspective is very interesting in order to explore the topology of toric varieties,
and it is ``easier'' because more explicit. One can easily visualize toric divisors and compute their intersections using the GLSM.
Nice reviews exist in the physics literature on this part of the story. See in particular:
\cite{Denef:2008wq}, \cite{Kreuzer:2006ax}, and the chapter 7 of \cite{Hori:2003ic}. 

\section*{Parting words}

Hundreds or thousands of physics papers on the gauge/gravity correspondence and on other subjects use some toric geometry tools.
It is my hope that these lectures will have helped other graduate students to contextualize this beautiful subject.

In particular, one cannot underestimate the power of algebraic geometry.
Hence we emphasized the algebraic concepts, in their most hands-on form, since this subject is not often known to graduate students of Physics.

\subsection*{Acknowledgements}
I thank the Gods of String Theory to have led me to this hellish world of theirs. I thank the organisers of the Modave Summer School 2008 for giving
me the opportunity to lecture on toric geometry in a relaxed environment. I would like to thank the participants of the schools for the warm atmosphere, and in particular I am grateful to A.~Bernamonti, J.~Evslin, S.~Kuperstein and V.~Wens for interesting discussions. 
I thank R.~Argurio and F.~Dehouck for proofreading and commenting on a previous version of these notes.
These lectures are to be published in the Modave Summer School Proceedings 2008.
C.C. is supported by the Belgian scholarship ``bourse de doctorat F.R.S. - FNRS''.

\appendix

\section{A few notions of algebra} \label{appendixAlgebra}
We just need a few definitions and propositions (without demonstration). For more details, see any algebraic geometry textbook, such as \cite{Griffiths}.

\paragraph{Ring.} A ring is a set $R$ equipped with two binary operations, $+$ and $\cdot$, such that\newline
(i) $(R,+)$ is a commutative group, \newline
(ii) $\cdot$ is associative and there exist a neutral element (called unity).  If moreover $\cdot$ is commutative we talk of a commutative ring (it is the case in these lectures).\newline
(iii) $\cdot$ is distributive over $+$ \, .\newline

Examples: The sel of all integers $\bbZ$ is a ring. Another example is the ring of polynomials in $n$ variables, denoted $\bbC[x_1,\cdots, x_n]$.

\paragraph{Ideal.} An ideal $I$ of a ring $R$ is a subset $I\subset R$ such that \newline
(i) $i,j\in I \, \Rightarrow \, i-j\in I$,\newline
(ii) $i\in I, r \in R \, \Rightarrow \, ir\in I$.

\paragraph{\textit{Proposition:}} $I$ an ideal of $R$ implies that the quotient $R/I$ is a ring too.

\paragraph{\textit{Notation:}} Given a set of elements $\{ r_1, \cdots, r_k \} \subset R$, we denote $(r_1,\cdots, r_k)$ the ideal generated by this set,
which is the smallest ideal of $R$ containing $\{ r_1, \cdots, r_k \} $.

\paragraph{Prime ideal.} An ideal $P\subset R$ is a prime ideal if for any ideals $I,J\subset R$,
\begin{displaymath}
I\cdot J = \{ij\in R \, |\, i\in I, j\in J \} \subset P \, \Rightarrow \, I\subset P \quad \mathrm{or}\quad  J \subset P.
\end{displaymath}

Exemple: In the ring $\bbC[x,y]$, the ideal $(xy)$ is not prime. It has a primary decomposition into $(x)$ and $(y)$.

\paragraph{Radical of an ideal.} Let $I$ be an ideal of $R$. The radical of $I$, denoted $\mathrm{rad}(I)$, is the intersection
of all the prime ideals containing $I$. ($\mathrm{rad}(I)$ is itself an ideal.) 

Example: In $\bbC[x,y]$, $\mathrm{rad}((x^ny^m))= (xy)$.\newline

An ideal $I\subset R$ is said to be radical if $\mathrm{rad}(I)=I$.

\paragraph{Height of a prime ideal.} The height $h(P)$ of a prime ideal $P\subset R$ is the largest integer $h$ such that there exist a chain of strict inclusions of prime ideals $P_i$
\be
P_0\subset P_1\subset \cdots \subset P_h=P\, .
\ee
It gives a notion of the dimension of an ideal. Moreover it can be shown that the dimension of
the affine variety corresponding to the quotient ring $\bbC[x_1,\cdots,x_n] /P$ is $n-h(P)$.

\paragraph{Zero divisor and integral domain.} An element $r\in R$, $r\neq 0$, is called a zero divisor if there exists $s\in R$, $s\neq 0$, such that $rs=0$. A commutative ring without zero divisor is called an integral domain.

\paragraph{\textit{Proposition:}} Given $R$ an integral domain, and $I\subset R$ an ideal, then $R/I$ is an integral domain if and only if $I$ is prime.

\subsection{Hilbert's Nullstellensatz}
Consider an ideal $I$ of $\bbC[x_1,\cdots,x_n]$. Given the algebraic subset  $Z(I)\subset \bbC^n$, as defined in section \ref{section Affine varieties}, is the knowledge of $Z(I)$ enough to reconstruct the ideal $I$? The answer is that you can only find $\mathrm{rad}(I)$. This is the content of the famous Hilbert's Nullstellensatz. More precisely:

\paragraph{Theorem.} For any ideal $I$ of $\bbC[x_1,\cdots, x_n]$,
\begin{displaymath}
\calJ(Z(I))=\mathrm{rad}(I),
\end{displaymath}
where $\calJ(Z(I))$ is the set of all polynomials vanishing on $Z(I)$.


\begin{thebibliography}{99}

\bibitem{Candelas:1988di}
  P.~Candelas, P.~S.~Green and T.~Hubsch,
  ``Finite Distances Between Distinct Calabi-Yau Vacua: (Other Worlds Are Just Around The Corner),''
  Phys.\ Rev.\ Lett.\  {\bf 62} (1989) 1956.

\bibitem{Candelas:1989js}
  P.~Candelas and X.~C.~de la Ossa,
  ``Comments on Conifolds,''
  Nucl.\ Phys.\  B {\bf 342} (1990) 246.
  
\bibitem{Candelas:1989ug}
  P.~Candelas, P.~S.~Green and T.~Hubsch,
  ``Rolling Among Calabi-Yau Vacua,''
  Nucl.\ Phys.\  B {\bf 330}, 49 (1990).
  

\bibitem{Argurio:2003ym}
  R.~Argurio, G.~Ferretti and R.~Heise,
  ``An introduction to supersymmetric gauge theories and matrix models,''
  Int.\ J.\ Mod.\ Phys.\  A {\bf 19} (2004) 2015
  [arXiv:hep-th/0311066].

\bibitem{Bilal:2001nv}
  A.~Bilal,
  ``Introduction to supersymmetry,''
  arXiv:hep-th/0101055.
  
\bibitem{Luty:1995sd}
  M.~A.~Luty and W.~Taylor,
  ``Varieties of vacua in classical supersymmetric gauge theories,''
  Phys.\ Rev.\  D {\bf 53} (1996) 3399
  [arXiv:hep-th/9506098].

\bibitem{Gray:2008yu}
  J.~Gray, A.~Hanany, Y.~H.~He, V.~Jejjala and N.~Mekareeya,
  ``SQCD: A Geometric Apercu,''
  JHEP {\bf 0805} (2008) 099
  [arXiv:0803.4257 [hep-th]].
    
\bibitem{Greene:1996cy}
  B.~R.~Greene,
  ``String theory on Calabi-Yau manifolds,''
  arXiv:hep-th/9702155.

\bibitem{Kreuzer:2006ax}
  M.~Kreuzer,
  ``Toric Geometry and Calabi-Yau Compactifications,''
  arXiv:hep-th/0612307.

\bibitem{Bouchard:2007ik}
  V.~Bouchard,
  ``Lectures on complex geometry, Calabi-Yau manifolds and toric geometry,''
  arXiv:hep-th/0702063.

\bibitem{Hubsch:1992nu}
  T.~Hubsch,
  ``Calabi-Yau manifolds: A Bestiary for physicists,''
{\it  Singapore, Singapore: World Scientific (1992) 362 p}


\bibitem{Denef:2008wq}
  F.~Denef,
  ``Les Houches Lectures on Constructing String Vacua,''
  arXiv:0803.1194 [hep-th].

    
\bibitem{joyce}
D.~Joyce,  ``Compact manifolds with special holonomy,''
 436 pages, Oxford Mathematical Monographs series, OUP, 2000.


\bibitem{Fulton}
W.~Fulton,  ``Introduction to Toric Varieties'', 
Princeton University Press, 1993.

\bibitem{Cox}
D.~Cox, ``Recent developments in toric geometry'',
 arXiv:alg-geom/9606016.


\bibitem{Skarke:1998yk}
  H.~Skarke,
  ``String dualities and toric geometry: An introduction,''
  arXiv:hep-th/9806059.

\bibitem{Sharpe:2003dr}
  E.~Sharpe,
  ``Lectures on D-branes and sheaves,''
  arXiv:hep-th/0307245.


\bibitem{Aspinwall:2004jr}
  P.~S.~Aspinwall,
  ``D-branes on Calabi-Yau manifolds,''
  arXiv:hep-th/0403166.



\bibitem{Griffiths}
P.~Griffiths, J.~Harris, ``Principles of Algebraic Geometry'', Wiley, 1978.


\bibitem{Aharony:1997bh}
  O.~Aharony, A.~Hanany and B.~Kol,
  ``Webs of (p,q) 5-branes, five dimensional field theories and grid
  diagrams,''
  JHEP {\bf 9801} (1998) 002
  [arXiv:hep-th/9710116].

\bibitem{Strominger:1995cz}
  A.~Strominger,
  ``Massless black holes and conifolds in string theory,''
  Nucl.\ Phys.\  B {\bf 451} (1995) 96
  [arXiv:hep-th/9504090].

\bibitem{Klebanov:2000hb}
  I.~R.~Klebanov and M.~J.~Strassler,
  ``Supergravity and a confining gauge theory: Duality cascades and
  chiSB-resolution of naked singularities,''
  JHEP {\bf 0008} (2000) 052
  [arXiv:hep-th/0007191].





\bibitem{Altmann1}
K.~Altmann, 
``The versal Deformation of an isolated toric Gorenstein Singularity,'' 
arXiv:alg-geom/9403004.




\bibitem{Pinansky}
  S.~Pinansky,
  ``Quantum deformations from toric geometry,''
  JHEP {\bf 0603} (2006) 055
  [arXiv:hep-th/0511027].


\bibitem{Argurio:2007vq}
  R.~Argurio and C.~Closset,
  ``A Quiver of Many Runaways,''
  JHEP {\bf 0709}, 080 (2007)
  [arXiv:0706.3991 [hep-th]].

\bibitem{Witten:1993yc}
  E.~Witten,
  ``Phases of N = 2 theories in two dimensions,''
  Nucl.\ Phys.\  B {\bf 403} (1993) 159
  [arXiv:hep-th/9301042].

\bibitem{Martelli:2004wu}
  D.~Martelli and J.~Sparks,
  ``Toric geometry, Sasaki-Einstein manifolds and a new infinite class of
  AdS/CFT duals,''
  Commun.\ Math.\ Phys.\  {\bf 262} (2006) 51
  [arXiv:hep-th/0411238].
  
\bibitem{Gauntlett:2004yd}
  J.~P.~Gauntlett, D.~Martelli, J.~Sparks and D.~Waldram,
  ``Sasaki-Einstein metrics on S(2) x S(3),''
  Adv.\ Theor.\ Math.\ Phys.\  {\bf 8} (2004) 711
  [arXiv:hep-th/0403002].

\bibitem{Hori:2003ic}
  K.~Hori {\it et al.},
  ``Mirror symmetry,''
{\it  Providence, USA: AMS (2003) 929 p}.


\end{thebibliography}
\end{document}